\documentclass[prl,twocolumn,epsf,psfig]{revtex4}

\def\widetext{\end{twocolumn}}
\input epsf
\begin{document}

\title{Collective oscillations of a Fermi gas near a Feshbach resonance}

\author{Theja N. De Silva\footnote{Email address: thejad@ccmr.cornell.edu}$^{a,b}$, 
Erich J. Mueller\footnote{Email address: emueller@ccmr.cornell.edu}$^{a}$}
\affiliation{$^{a}$ Laboratory of Atomic and Solid State Physics, Cornell University, Ithaca, New York 14853, USA. \\
$^{b}$ Department of physics, University of Ruhuna, Matara, Sri Lanka.}

\begin{abstract}
A sum rule approach is used to calculate the zero temperature oscillation frequencies of a two component trapped atomic 
Fermi gas in the BCS-Bose Einstein condensation crossover 
region. These sum rules are evaluated using a local density approximation which explicitly includes Feshbach molecules. 
Breathing modes show non-monotonic behavior as a function of the interaction strength, while quadrupole modes are insensitive to 
interactions for both spherically symmetric and axially symmetric traps. 
Quantitative agreement is found with experiments on atomic $^6Li$ system and with other theoretical approaches.

\end{abstract}

\maketitle

\section{I. Introduction}

The successful trapping of fermionic atomic gases in magnetic and optical traps and their subsequent cooling to 
quantum degeneracy has attracted much interest. Experimentalists wield large degrees of experimental control over 
these low temperature and low density gases, 
including the ability to manipulate the effective atom-atom interaction by tuning an external magnetic 
field~\cite{greiner,jochim,zwier,regal}. The 
Zeeman shift from the magnetic field adjusts the energy of a two atom bound state relative to the scattering continuum.
Around the Feshbach (FB) resonance, where the binding energy is exactly equal to 
the energy of colliding atoms, a small change in magnetic field can have a dramatic effect 
on the effective interaction~\cite{sl}. On one side of the resonance (above the resonance) the atomic gas has 
attractive interactions, while on the other side (below the resonance) effective interactions are repulsive. At 
resonance, the scattering length of the atoms ($a$) is divergent and one expects universal behavior~\cite{uni}. If the temperature is low 
enough, then fermionic atoms can pair and undergo Bardeen, Cooper and Schrieffer (BCS) condensation on the attractive 
side (BCS regime) of the resonance. On the repulsive side (BEC regime), there exists Bose molecules (FB molecules)
which are short-range fermionic atom pairs and these molecules can undergo Bose-Einstein condensation (BEC) at low 
temperature. By sweeping the external magnetic field through the FB resonance, one expects a continuous crossover between these 
limits~\cite{laggett,noz,mod2,mod3,cot,th,cre}.

Here we study the collective oscillations of a Fermi gas in the crossover region. 
As a readily accessible dynamical observable, collective oscillations of a trapped gas can yield crucial information on the state of 
the system. Recently, both transverse and axial breathing modes of a superfluid $^6Li$ atomic system have been measured in a highly 
anisotropic trap~\cite{kinast1,barten1,kinast2,kinast3}. Several theoretical attempts have been made to understand 
these collective oscillations~\cite{stringari1,hu,heisel,kim,manini,astrak,aurel,ohashi1}. Stringari~\cite{stringari1} used 
a hydrodynamic theory to calculate the collective oscillations far from resonance $(|a| \ll 1)$ and at unitarity 
($|a| \rightarrow \infty$). By interpolating from these results he predicted that the frequency of 
the transverse breathing mode in a highly elongated trap should exhibit a non-trivial dependence on the scattering 
length~\cite{stringari1}. This prediction was further supported by studies based on semi-empirical forms of the equation of 
state, time dependent density functional approaches, and asymptotic expansions~\cite{hu, heisel, kim,astrak,aurel,manini}. Additionally, 
with explicit inclusion of Feshbach molecules, Ohashi et al~\cite{ohashi1} used a Hartree-Fock-Bogoliubov (HFB) approach to study the 
collective oscillation in the BCS-BEC crossover region. While their theory predicts correct results in the weak coupling BCS limit, 
their results in the weakly repulsive BEC limit are those of a non-interacting (rather than interacting) Bose condensate.

Unlike some of these previous studies, our approach, based upon the ``compressibility sum rule'' and the 
``f-sum rule'', describes this system in all regions of the phase diagram. It is more flexible than the methods based 
upon isotropic indices, explicitly includes Feshbach molecules, and correctly predicts all limiting cases. 

The paper is organized as follows. 
In section II, the two channel coupled fermion-boson model is introduced. Following the pioneering work in the context of 
superconductivity by Eagles~\cite{eagles}, Leggett~\cite{laggett}, 
Nozieres and Schmitt-Rink~\cite{noz}, 
we simplify this model by using a mean field theory described in section III. In section IV, we use sum rules to derive the 
collective breathing modes 
and quadrupole modes for both spherically symmetric and axially symmetric traps. Finally, 
we discuss our results with comparison to the experimental and other theoretical predictions in section V.

\section{II. The Model}

We consider a dilute gas of interacting Fermi atoms in two hyperfine states trapped in an external potential $V(\vec{r})$. 
Since the atomic 
gas is cold and very dilute, the inter-atomic distance is 
large compared to the range of the atomic potential and s-wave scattering dominates over all other scattering 
channels. Therefore, we neglect interactions between fermions in the same hyperfine 
states and approximate the attractive inter-atomic potential between fermions in different hyperfine states by a contact potential 
$-u_{\rm{bare}} \delta(\vec{r}\prime-\vec{r})$. Furthermore, we take into account coupling between molecular bosons associated 
with the 
Feshbach resonance and Fermi atoms by a local conversion term, proportional to 
$g_{\rm{bare}}$. The Feshbach resonance is achieved by tuning the threshold 
energy $2 \nu_{\rm{bare}}$, which varies linearly with magnetic field. The atomic gas is then described by 
the Hamiltonian~\cite{mod2, mod3,th,mod1}

\begin{eqnarray}
H = \sum_{\sigma}\int d^3\vec{r} \psi_{\sigma}^{\dagger}(\vec{r})\biggr(-\frac{\hbar^2\nabla^2}{2m}-\mu_0+V(\vec{r})\biggr)
\psi_{\sigma}(\vec{r}) \nonumber \\
+\int d^3\vec{r} \phi^{\dagger}(\vec{r})\biggr(-\frac{\hbar^2\nabla^2}{2M}+2\nu_{\rm{bare}}-2\mu_0+2V(\vec{r})\biggr)\phi(\vec{r}) \nonumber \\
-u_{\rm{bare}}\int d^3\vec{r}\psi^{\dagger}_{\uparrow}(\vec{r})\psi^{\dagger}_{\downarrow}(\vec{r})\psi_{\downarrow}(\vec{r})
\psi_{\uparrow}(\vec{r}) \nonumber \\
+g_{\rm{bare}}\int d^3\vec{r}\biggr(\phi^{\dagger}(\vec{r})\psi_{\downarrow}(\vec{r})\psi_{\uparrow}(\vec{r})+h.c\biggr).
\end{eqnarray}

\noindent where $m (M=2m)$ is the mass of the Fermi atoms (Bose molecules), and $\mu_0 (\mu_M=2 \mu_0)$ is the Fermi (Bose) 
chemical potential. The field operators $\psi_{\sigma}(\vec{r})[\phi(\vec{r})]$ obey the usual fermionic (bosonic) 
anticommutation (commutation) rules, and describe the annihilation of a fermion (boson) at position $\vec{r}$ in the 
hyperfine state $\sigma = \uparrow$ or $\downarrow$. In general, the trapping potential for Fermi atoms is taken as 
$V(\vec{r})=\sum_{\alpha \beta} \frac{m \omega^2_{\alpha \beta}}{2} r_{\alpha} r_{\beta}$. For Bose molecules, the trapping potential 
is $2V(\vec{r})$. Notice, the trapping frequencies $\omega_{\alpha \beta}$, $\alpha,\beta = x,y,z$ are the same 
for both species of fermions and molecular bosons. In this paper, we consider only spherical symmetric and axially 
symmetric traps. For the case of a spherically symmetric trap, $\omega_{\alpha \beta}=\omega \delta_{\alpha,\beta}$, 
where $\delta$ is the Kronecker delta. For 
axially symmetric traps, $\omega_{\alpha \beta}=\omega_{\perp}\delta_{\alpha,\beta}$ for $\alpha, \beta=x,y$ and 
$\omega_{\alpha \beta}=\omega_{z}\delta_{\alpha,\beta}$ for $\alpha, \beta=z$.

\section{III. Mean field equations}

The gap equation and the number equation are obtained by using a mean field approximation. Introducing two local mean fields for the BCS 
condensate and BEC condensate respectively,  
$\Delta(\vec{r})=u_{\rm{bare}}\langle \psi_{\downarrow}(\vec{r})\psi_{\uparrow}(\vec{r})\rangle$ and 
$\phi_{m}(\vec{r})=\langle \phi(\vec{r})\rangle$, 
the mean field Hamiltonian is written as

\begin{eqnarray}
H_{MF} = \sum_{\sigma}\int d^3\vec{r} \psi_{\sigma}^{\dagger}(\vec{r})\biggr(-\frac{\hbar^2\nabla^2}{2m}-\mu(\vec{r})\biggr)\psi_{\sigma}(\vec{r}) \nonumber \\
+\int d^3\vec{r} \phi^{\dagger}(\vec{r})\biggr(-\frac{\hbar^2\nabla^2}{2M}+2\nu_{\rm{bare}}-2\mu(\vec{r})\biggr)\phi(\vec{r}) \nonumber \\
-\int d^3\vec{r}\tilde{\Delta}(\vec{r}) \biggr(\psi^{\dagger}_{\uparrow}(\vec{r})\psi^{\dagger}_{\downarrow}(\vec{r})+h.c\biggr).
\end{eqnarray}

\noindent where we defined a local composite order parameter 
$\tilde{\Delta}(\vec{r}) = \Delta(\vec{r})-g_{\rm{bare}}\phi_m(\vec{r}) 
= u_{\rm{bare}}^{\rm{eff}}\langle \psi_{\downarrow}(\vec{r})\psi_{\uparrow}(\vec{r})\rangle$ 
with $u_{\rm{bare}}^{\rm{eff}}(\vec{r}) = u_{\rm{bare}}-\frac{g_{\rm{bare}}^2}{2 \mu(\vec{r})-2 \nu_{\rm{bare}}}$. Where 
$\langle A \rangle$ represents the expectation 
value of the operator $A$. The local chemical potential 
$\mu (\vec{r}) = \mu_0-V(\vec{r})$ is to be treated with 
local density approximation (LDA). Within the LDA, we assume that the density varies so slowly so that it can be treated as locally 
uniform. Therefore, we determine the solutions of the equations at each point in space as for a homogeneous system. 
This approximation is reasonable for the current experiments as the number of atoms in the trap is very large~\cite{lda1,lda2}. 
Also, we assume that at zero temperature all the molecules are Bose-condensed. 
The two order parameters introduced before are not independent and they are strongly coupled to each other by

\begin{eqnarray}
\frac{g_{\rm{bare}} \Delta}{u_{\rm{bare}}}+(2 \nu_{\rm{bare}}-2 \mu)\phi_m = 0.
\end{eqnarray}

\noindent We have suppressed the position dependence of the local quantities in this equation and the following. 
Notice, as a result of this strong coupling between $\Delta(\vec{r})$ and $\phi_m(\vec{r})$, both BCS and BEC 
condensates are non-zero through out the entire BCS-BEC crossover region. 

Diagonalizing the mean field Hamiltonian by a Bogoliubov transformation, the usual BCS gap equation and number equation at zero 
temperature are given by

\begin{eqnarray}
1 = u_{\rm{bare}}^{\rm{eff}}(\vec{r}) \int \frac{d^3\vec{k}}{(2 \pi)^3}\frac{1}{2E_k}
\end{eqnarray}

\begin{eqnarray}
n_F(\vec{r}) = \int \frac{d^3\vec{k}}{(2 \pi)^3}(1-\frac{\epsilon_k-\mu}{E_k})
\end{eqnarray}

\noindent Here $\epsilon_k=\hbar^2k^2/2m$ is the kinetic energy, $E_k = \sqrt{(\epsilon_k-\mu)^2+\tilde{\Delta}^2}$ is 
the BCS-Bogoliubov excitation energy, and $n_F(\vec{r})$ is the density of the Fermi atoms. The total density is the sum of 
fermionic atom density and Bose molecular density, 
$n(\vec{r}) = n_F(\vec{r})+2 \phi_{m}(\vec{r})^2$. Because of the short range nature of the interaction, the gap equation 
shows an ultra-violet divergence. Using a standard regularization procedure~\cite{mod2, mod3,cre,reg} to subtract off the 
divergences, a renormalized effective interaction $u_{\rm{eff}}(\vec{r})$ is introduced through the relation 
$\frac{1}{u_{\rm{eff}}(\vec{r})} = \frac{1}{u_{\rm{bare}}^{\rm{eff}}(\vec{r})} - \int \frac{d^3\vec{k}}{(2 \pi)^3}\frac{1}{2\epsilon_k}$. 
Then the renormalized BCS gap equation is written as

\begin{eqnarray}
1 = u_{\rm{eff}}(\vec{r}) \int \frac{d^3\vec{k}}{(2 \pi)^3}(\frac{1}{2E_k}-\frac{1}{2 \epsilon_k})
\end{eqnarray}

\noindent where renormalized effective interaction is written in terms of renormalized pairing interaction  $u$, 
renormalized FB resonant strength $g$ and renormalized detuning $\nu$ as $u_{\rm{eff}}(\vec{r}) 
= u-\frac{g^2}{2 \mu(\vec{r})-2 \nu} = -\frac{4 \pi \hbar^2 a(\vec{r})}{m}$.
Notice, we have introduced a position dependent scattering length $a(\vec{r})$ which crucially depends on the local chemical 
potential $\mu(\vec{r})$ and the threshold energy $2\nu$ (See FIG. 1 
below). The composite order parameter $\tilde{\Delta}(\vec{r})$ 
is now written in terms of renormalized $g$ as $\tilde{\Delta}(\vec{r}) = \Delta(\vec{r}) - g\phi_m(\vec{r})$. 
The renormalized non-resonant pairing interaction $u$ and 
renormalized FB resonant strength $g$ must be extracted from 
experiments. Assuming the atom scattering length near FB resonance takes the form $a_s = a_b(1-\frac{W}{B-B_0})$, with $a_b$ the 
background scattering length, $B_0$, the resonant magnetic field, $W$, the resonance width, they are given by 
$u = \frac{4\pi\hbar^2a_b}{m}$ and $g = \sqrt{\frac{4\pi\hbar^2\mu_{\delta}W|a_b|}{m}}$, where $\mu_{\delta}$ is the difference of the 
atomic magnetic moments between the closed and the open scattering channels. From experiments 
on $^6Li$ atomic system by Bartenstein at el~\cite{barten2}, we see that $u \sim 9.0 \times 10^7k_B$ nm$^3$$\mu$K and 
$g \sim 1.7 \times 10^6k_B$ nm$^{3/2}$$\mu$K, corresponding to 
the broad resonance of $^6Li$ system at magnetic field $B_0 \sim 837$G. In addition to this broad resonance, 
this system shows a narrow resonance at $B_0 \sim 545$G~\cite{nr} which 
gives $g \sim 2.4 \times 10^3k_B$ nm$^{3/2}$$\mu$K. 
For the broad resonance, the crossover occurs at 
$|\nu| \gg \epsilon_f$, while at the narrow resonance, it occurs at $|\nu| \sim \epsilon_f$, 
where $\epsilon_f = \frac{\hbar^2k_f^2}{2m}$ is the Fermi energy. However, when 
viewed as a function of scattering length, one finds the same crossover physics irrespective of the width of 
the resonance~\cite{ohashi2}. In this paper, we have done our calculation for the broad resonance and present 
our results as a function of the dimensionless parameter $(k_fa_s)^{-1}$, 
where $a_s$ is the scattering length at the center of the trap defined through the relation, 
$u-\frac{g^2}{2\mu_0-2\nu} = -\frac{4\pi \hbar^2a_s}{m}$, 
hence our results must be applicable to the narrow resonance as well.

Eq. (5) and Eq. (6) are to be solved with the constraint that the total number 
of atomic particles in the trap is $N = \int d^3\vec{r} n(\vec{r})$. The size of the cloud $r_0$ is determined by the condition 
$n(\vec{r}=\vec{r}_0) = 0$. Depending on the sign of the scattering length, this condition gives two 
different formulas for determining the size of the cloud.   
For $a(\vec{r}) < 0$ (BCS regime), the boundary is given by $\mu(\vec{r}=\vec{r}_0)=0$
and, for $a(\vec{r}) > 0$ (BEC regime), it is given by 
$\mu(\vec{r}=\vec{r}_0) = -\frac{\hbar^2}{ma(\vec{r_0})^2}$.
In the weak coupling limit ($k_fa \rightarrow 0^-$ or equivalently $\nu \rightarrow +\infty$), Eq. (5) and Eq. (6) 
reduce to the standard 
BCS results, $n(\vec{r}) \propto \mu^{3/2}$ and $\mu \sim \epsilon_f$. When  $k_fa \rightarrow 0^+$ (or equivalently 
$\nu \rightarrow -\infty$), 
they lead to a Bose gas with 
$n(\vec{r}) \propto |\mu|^ {2 } (\frac {\hbar } {a  \sqrt {2m|\mu|}}-1)$ and $\mu \sim - \frac {\hbar^2} {2 m a^2 }$. 
At unitarity $(a=\pm \infty)$, physical quantities of the system must be independent of the scattering length. 
Therefore, the system shows universal behavior with $n(\vec{r})  \propto  \mu^ {3/2 }$~\cite{uni}. 

We plot the spatial variation of the inverse scattering length and density in FIG. 1 in a spherical trap, using parameters from 
the two resonance in $^6Li$. We choose $\nu \sim 4\epsilon_f$ for the narrow resonance and $\nu \sim 10^4 \epsilon_f$ 
for the broad resonance so that $a(0)k_f$ is the same in each case. The large variation in $a(\vec{r})$ for the narrow resonance 
is due to the fact that $\nu \sim \epsilon_f$.

\begin{figure}[tbh]
\epsfxsize=8.0cm 
\centerline{\epsffile{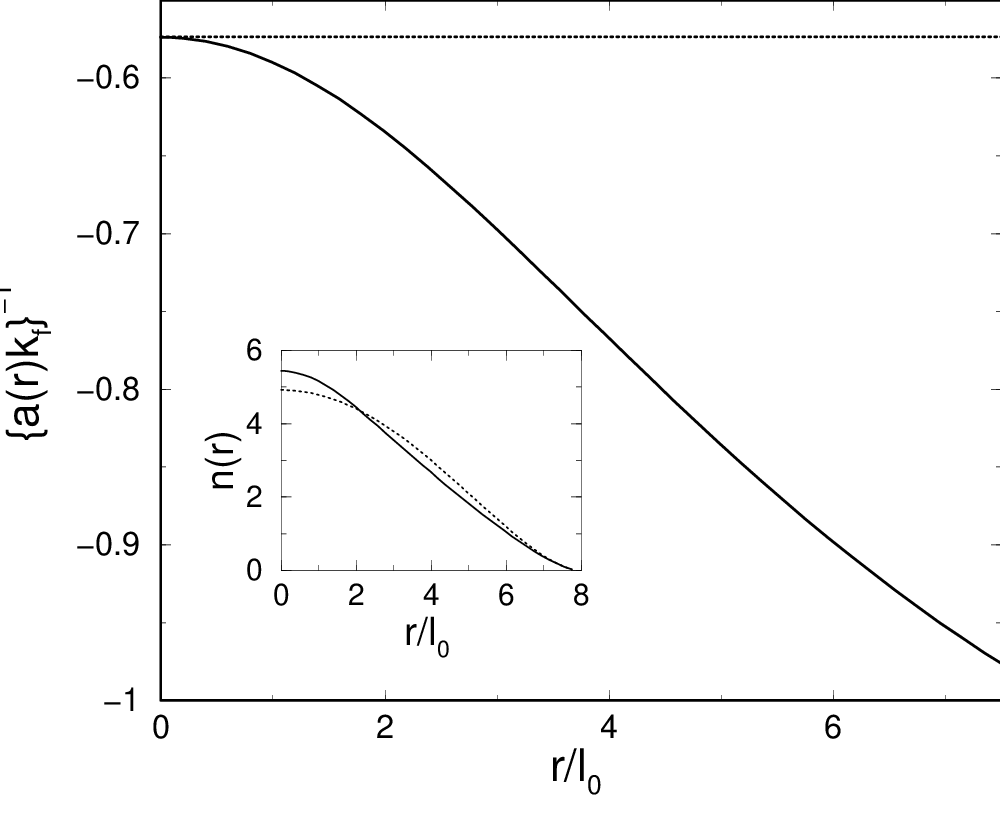}}
\caption{ Spatially dependent inverse scattering length $\{k_fa(r)\}^{-1}$ for spherically symmetric trap as a function 
of $r/l_0$, where $l_0$ and $k_f$ are oscillatory length and Fermi wave vector at the center of the trap respectively. 
Dotted line represents the broad resonance while solid line represents 
the narrow resonance. Inset shows the scaled density variation at the same parameter regime.}
\end{figure}

\section{IV. Sum rules and the collective oscillations}
 
We use a sum rule approach to study the propagation of collective modes in harmonically trapped Fermi gas in the presence of 
FB resonance. Sum rules are derived from the linear response theory and allow us to compute rigorous upper bounds to the energies of
 the collective oscillations at zero temperature. The response function 
$\chi(z)$ at complex frequency $z$ is written in terms 
of the imaginary part of the response function $\chi ^{\prime \prime}(\omega)$ as

\begin{eqnarray}
\chi(z) = \int \frac{d \omega}{\pi} \frac{\chi ^{\prime \prime}(\omega)}{(\omega-z)}.
\end{eqnarray}
 
\noindent The explicit determination of the dissipative 
component of the response function $\chi^{\prime \prime}(\omega)$ 
requires the full solution of the Hamiltonian. However, useful information on the behavior of the system can be derived by 
using the method of sum rules. Expanding the response function in terms of $1/z$, moment expansion of $\chi(z)$ is given by

\begin{eqnarray}
\chi(z)=\sum_p \frac{1}{z^p} m_{p-1}. 
\end{eqnarray}

\noindent where the $p$th moment is defined as 

\begin{eqnarray}
m_p = \int \frac{d \omega}{\pi} \omega^p \chi ^{\prime \prime}(\omega).
 \end{eqnarray}

\noindent The collective oscillation 
of the system is then bounded by $\sqrt{\frac{m_p}{m_{p-2}}}$~\cite{bohi}. With the sum rule approach, 
it is possible to compute these moments 
without evaluating response function. Considering quadrupole  excitations 
$Q_{\alpha \beta}=\int d^3\vec{r}r_{\alpha} r_{\beta}(\sum_{\sigma}\psi_{\sigma}^{\dagger}\psi_{\sigma}+2\phi^{\dagger}\phi)$ 
and the 
perturbed Hamiltonian 
$H_{pert} = \int \sum_{\alpha \beta}d^3\vec{r}\lambda_{\alpha \beta}r_{\alpha} r_{\beta}(\sum_{\sigma}\psi_{\sigma}^{\dagger}\psi_{\sigma}+2\phi^{\dagger}\phi)$, 
the response function is written as 
$\chi^{\alpha \beta}_{\gamma \delta} = \frac{\partial Q_{\alpha \beta}}{\partial \lambda_{\gamma \delta}}$, 
where $\lambda_{\alpha \beta} = \frac{1}{2}m\omega_{\alpha \beta}^2$ and $\alpha$, $\beta$ run over the Cartesian indeces $x$, $y$ 
and $z$. The moments, 
$(m_{-1})^ {\alpha \beta}_{ \gamma \delta}$ and $(m_1)^ {\alpha \beta}_{ \gamma \delta}$ are given by
 
\begin{eqnarray}
(m_{-1})^ {\alpha \beta}_{ \gamma \delta}=\int \frac{d \omega}{2 \pi \omega} Im (\chi^ {\alpha \beta}_{ \gamma \delta}) \nonumber \\
= -\chi^ {\alpha \beta}_{ \gamma \delta}(\omega=0)
=-\frac{1}{m\omega_{\gamma \delta}}\frac{\partial \langle Q_{\alpha \beta} \rangle}{\partial \omega_{\gamma \delta}}
\end{eqnarray}

\noindent and
 
\begin{eqnarray}
(m_1)^ {\alpha \beta}_{ \gamma \delta} = \int \frac{d \omega}{2 \pi} \omega Im (\chi^ {\alpha \beta}_{ \gamma \delta})
= \langle[[Q_{\alpha \beta},H],Q_ {\gamma \delta} ]\rangle \nonumber \\
 = \frac{1}{m}(\delta_{\alpha \gamma} \langle Q_ {\beta \delta}\rangle+ \delta_{\alpha \delta} \langle Q_ {\beta \gamma}\rangle+ \delta_{\beta \delta} \langle Q_ {\alpha \gamma}\rangle+ \delta_{\beta \gamma} \langle Q_ {\alpha \delta}\rangle).
\end{eqnarray}

\noindent where $[A,B]$ represents the commutator between the operators $A$ and $B$. The squares of the collective 
mode frequencies are approximated by the eigenvalues of the matrix $\tilde{M} = (m_{-1})^{-1} (m_1)$.

\subsection{Spherically symmetric trap}

In the case of a spherically symmetric trap, $V(\vec{r})=\frac{1}{2}m\omega^2r^2=br^2$. Solving for the eigenvalues of the 
matrix $\tilde{M}$, the breathing mode $\omega_m$, and the quadrupole mode $\omega_{q}$ associated 
with eigenvectors $\{1,1,1\}$ and $\{-1,1,0\}$ are, respectively, given by

\begin{eqnarray}
\omega_m^2 = -\frac{4}{m} \frac{\langle r^2 \rangle}{\frac{\partial \langle r^2 \rangle}{\partial b}}
\end{eqnarray}

\noindent and 

\begin{eqnarray}
\omega_q^2 = -\frac{4}{m} \frac{\langle r^2 \rangle}{3}\frac{1}{\frac{\partial \langle x^2 \rangle}{\partial b_x}-\frac{\partial \langle x^2 \rangle}{\partial b_y} } 
\nonumber \\
= -\frac{4}{m} \frac{\langle r^2 \rangle}{3}\biggr( \frac{1}{-\frac{1}{2}\int d^3\vec{r}(x^2-y^2)^2 \frac{\partial n(\vec{r})}{\partial \mu}}\biggr)
\end{eqnarray}

\noindent where, $\langle r^2 \rangle = \int d^3\vec{r} r^2 n(\vec{r})$ is the expectation value of the 
operator $\vec{r}^2$ with the total 
atomic density $n(\vec{r})$. In the weak coupling BCS limit, the number of molecular bosons can be neglected, hence one can show
that $\mu_0 \propto \sqrt{b}$ and $\langle r^2 \rangle \propto b^{-\frac{1}{2}}$. As a result, the breathing mode and the 
quadrupole mode 
approach to the expected results $\omega_m \rightarrow 2 \omega$ 
and $\omega_q \rightarrow \sqrt{2}\omega$ as in the case of a non interacting Fermi gas. In the opposite limit, 
the number of Fermi atoms can be neglected, so that one can show that 
$\mu_0^{\frac{7}{2}}(\frac{\hbar}{a \sqrt{2m\mu_0}}-1)(\frac{\hbar^2}{2ma^2\mu_0}-1)^{\frac{3}{2}} \propto b^{\frac{3}{2}}$ and 
$\langle r^2 \rangle \propto \mu_0b^{-1}(\frac{\hbar^2}{2ma^2\mu_0}-1)$ which resulted in $\omega_m \rightarrow \sqrt{5}\omega$ and 
$\omega_q \rightarrow \sqrt{2}\omega$. Notice, in this limit, the mode frequencies are identical to the result obtained by 
the Thomas-Fermi limit of a Bose gas~\cite{gp}. As opposed to the HFB theory~\cite{ohashi1}, our approach retains effective 
molecular interaction at the weak repulsive BEC limit and recovers 
hydrodynamic results for both limits~\cite{gp,vic}. Also, at unitarity, the chemical potential dependence on the density is the same as 
that of a non interacting Fermi gas, hence the mode frequencies are expected to be the same as those of non interacting 
limit of the Fermi gas. We numerically evaluate Eq. (12) and Eq. (13) in the LDA. The breathing 
mode frequencies in the crossover region are plotted in FIG. 2 as a function of dimensionless parameter 
$(k_fa_s)^{-1}$. In addition to finding the expected results as $(k_fa_s)^{-1} \rightarrow \pm \infty$, we 
find a dip near unitarity on the BCS side. 

Our numerical calculation confirms that the quadrupole mode calculated from 
Eq. (13) is independent of the interaction and 
the aspect ratio of the trap (see below), and its frequency is $\sqrt{2}\omega$ through out the entire BCS-BEC crossover.

\begin{figure}[tbh]
\epsfxsize=8.0cm 
\centerline{\epsffile{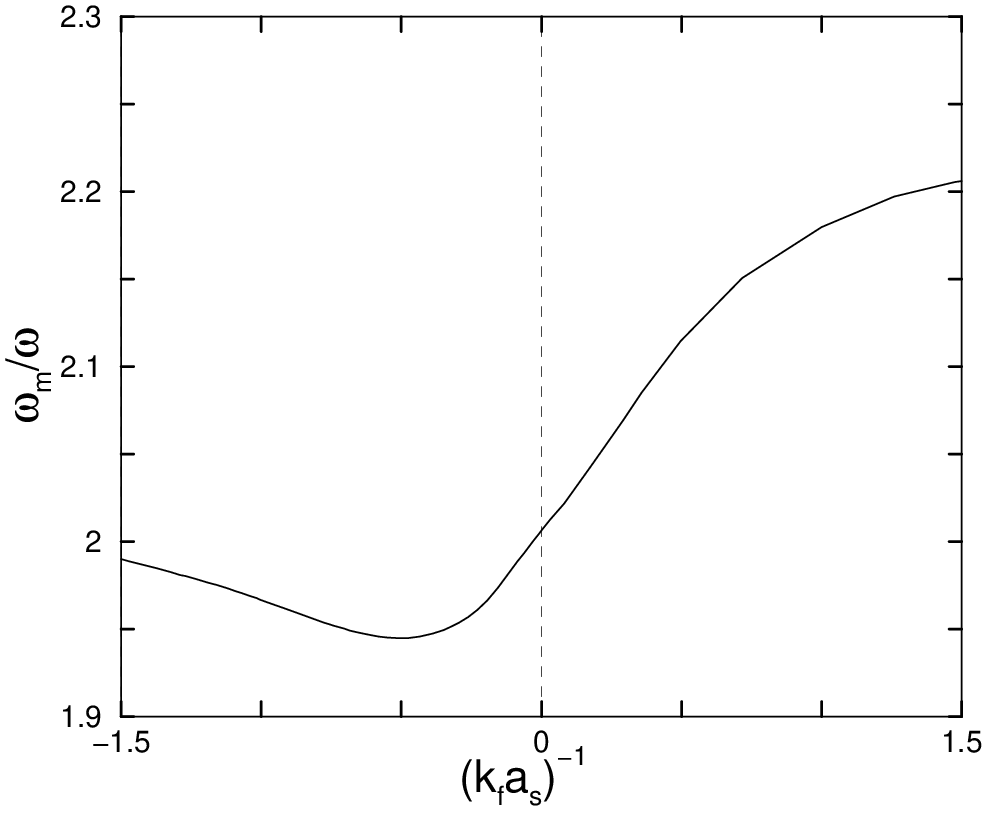}}
\caption{Zero temperature breathing mode frequency of a two component Fermi gas in a spherically symmetric 
trap within the BCS-BEC crossover 
region. The frequency is given as a function of a dimensionless parameter $(k_fa_s)^{-1}$. Our theory 
predicts $\omega_m/\omega \rightarrow 2$ 
at the weak coupling BCS limit $[(k_fa_s)^{-1} \rightarrow -\infty]$ and $\omega_m/\omega \rightarrow \sqrt{5}$ at the 
weakly repulsive BEC limit $[(k_fa_s)^{-1} \rightarrow +\infty]$ as one 
expects from hydrodynamic 
theory. At unitarity, $\omega_m/\omega \rightarrow 2$ as required by universality.}
\end{figure}

\subsection{Axially symmetric trap}

In the case of axially symmetric trap, $V(\vec{r})= \frac{1}{2}m\omega_{\perp}^2(\rho^2+\lambda z^2)=b_{\perp}\rho^2+b_zz^2$, where 
$\rho^2=x^2+y^2$. For this case, solving for the eigenvalues of the matrix $\tilde{M}$, the breathing modes $\omega_{\pm}$ are given by

\begin{eqnarray}
\omega_{\pm}^2 = -\frac{2}{m}\frac{1}{(Q_1Q_2-2 Q_3)}\biggr (\frac{1}{2}\langle \rho^2 \rangle Q_2+\langle z^2 \rangle Q_1 
\nonumber  \\
\pm \{\frac{1}{4}{\langle \rho^2\rangle}^2Q_2^2 +{\langle z^2\rangle}^2Q_1^2 \nonumber \\
+ \frac{1}{2}\langle \rho^2\rangle\langle z^2 \rangle (8Q_3-2Q_1Q_2)\}^{\frac{1}{2}}\biggr ). 
\end{eqnarray}

\noindent where we define, $Q_1 = \frac{1}{2} \frac{\partial  \langle \rho^2\rangle}{\partial b_{\perp}}$, 
$Q_2 = \frac{\partial \langle z^2\rangle}{\partial b_z}$ and 
$Q_3 = \frac{1}{4} \frac{ \partial \langle \rho^2 \rangle}{\partial b_z} \frac{\partial \langle z^2 \rangle}{\partial b_{\perp}}$. 
The quadrupole 
mode $\omega_q$ is given by  

\begin{eqnarray}
\omega_q^2 = -\frac{4}{m} \frac{\langle \rho^2 \rangle}{2} \biggr(\frac{1}{-\frac{1}{2}\int d^3\vec{r}(x^2-y^2)^2 \frac{\partial n(\vec{r})}{\partial \mu}}\biggr).
\end{eqnarray}

\noindent The expectation values $\langle .... \rangle$ are defined as before. In the weak coupling BCS limit, 
the trapping frequencies dependence on the chemical potential as $\mu_0 \propto b_{\perp}b^{\frac{1}{2}}_z$. 
At this limit, the frequencies dependence on the expectation values of the operators $\rho^2$ and $z^2$ are 
$\langle \rho^2 \rangle \propto \mu_0^4b_{\perp}^{-2}b_z^{-\frac{1}{2}}$ 
and $\langle z^2 \rangle \propto \mu_0^4b_{\perp}^{-1}b_z^{-\frac{3}{2}}$. Then Eq. (13) yields $\omega^2_{\pm}\rightarrow 
4\omega_{\perp}^2(\frac{5}{12}+\frac{\lambda}{3}\pm \sqrt{\frac{25}{144}+\frac{\lambda^2}{9}-\frac{2\lambda}{9}})$. 
For highly the elongated traps ($\lambda \ll 1$) in which most of the 
experiments are carried out, $\omega_{+} \rightarrow \sqrt{\frac{10}{3}}\omega_{\perp}$ and 
$\omega_{-} \rightarrow \sqrt{\frac{12}{5}}\omega_{z}$. In the opposite limit (weakly repulsive BEC limit), the trapping frequencies and 
the chemical potential are related by the expression, 
$|\mu_0|^\frac{7}{2}(\frac{\hbar}{a\sqrt{2m|\mu_0|}}-1)(\frac{\hbar^2}{2ma^2|\mu_o|}-1)^{\frac{3}{2}} 
\propto b_{\perp}b_z^{\frac{1}{2}}$. In this weakly repulsive BEC limit, the expectation values of the operators $\rho^2$ and $z^2$, 
the chemical potential 
and, the trapping frequencies are related as $\langle \rho^2 \rangle \propto |\mu_0|^\frac{9}{2}(\frac{\hbar}{a\sqrt{2m|\mu_0|}}-1)
(\frac{\hbar^2}{2ma^2|\mu_o|}-1)^{\frac{5}{2}}b_{\perp}^{-2}b_z^{-\frac{1}{2}}$ and $\langle z^2 \rangle \propto 
 |\mu_0|^\frac{9}{2}(\frac{\hbar}{a\sqrt{2m|\mu_0|}}-1)
(\frac{\hbar^2}{2ma^2|\mu_o|}-1)^{\frac{5}{2}}b_{\perp}^{-1}b_z^{-\frac{3}{2}}$. The Eq. (14) then yields $\omega^2_{\pm}\rightarrow 
5\omega_{\perp}^2(\frac{2}{5}+\frac{3\lambda}{10}\pm \sqrt{\frac{4}{25}+\frac{9\lambda^2}{100}-\frac{4\lambda}{25}})$. For the case of 
highly elongated traps,  $\omega_{+} \rightarrow 2\omega_{\perp}$ and 
$\omega_{-} \rightarrow \sqrt{\frac{5}{2}}\omega_{z}$. At unitarity, collective frequencies are expected to be the same as 
those at the weak coupling BCS limit. The mode frequencies $\omega_{+}$ and $\omega_{-}$ in the crossover region are 
presented in FIG. 3 and FIG. 4 for the case of $\lambda = 0.001$. The behavior of the modes are the same as that of a spherically 
symmetric trap, agreeing with 
the hydrodynamic theory of a non interacting Fermi gas and a large N Bose gas at the asymptotic limits, while 
displaying a dip near the unitary limit on BCS side. A similar 
behavior of the collective oscillations has been recovered using either a scaling ansatz based on a mean field description 
of the BCS-BEC crossover~\cite{hu} or the generalized 
Hylleraas-Undheim method~\cite{kim}. As suggested by Hu et al~\cite{hu}, these 
dips in the collective frequencies may be attributed to the fact that pairing enhances the compressibility of the atomic gas. 
The measured axial breathing modes have quantitative agreements in the BEC side with experiments carried out by 
Bartenstein et al~\cite{barten1}. The measured transverse breathing modes are in 
quantitative agreements with the experiment carried out by Kinast et al~\cite{kinast1}. For comparison, we plotted experimental 
results from Ref. ~\cite{kinast1} and Ref. ~\cite{barten1} as solid circles in FIG. 3 and FIG. 4. In section V, these 
experimental results are compared with our 
prediction in more details.

As in the case of a spherically symmetric trap, the quadrupole 
mode $\omega_q \rightarrow \sqrt{2}\omega_{\perp}$ and our 
calculation confirms that it is independent of the interaction.

\begin{figure}[tbh]
\epsfxsize=8.0cm 
\centerline{\epsffile{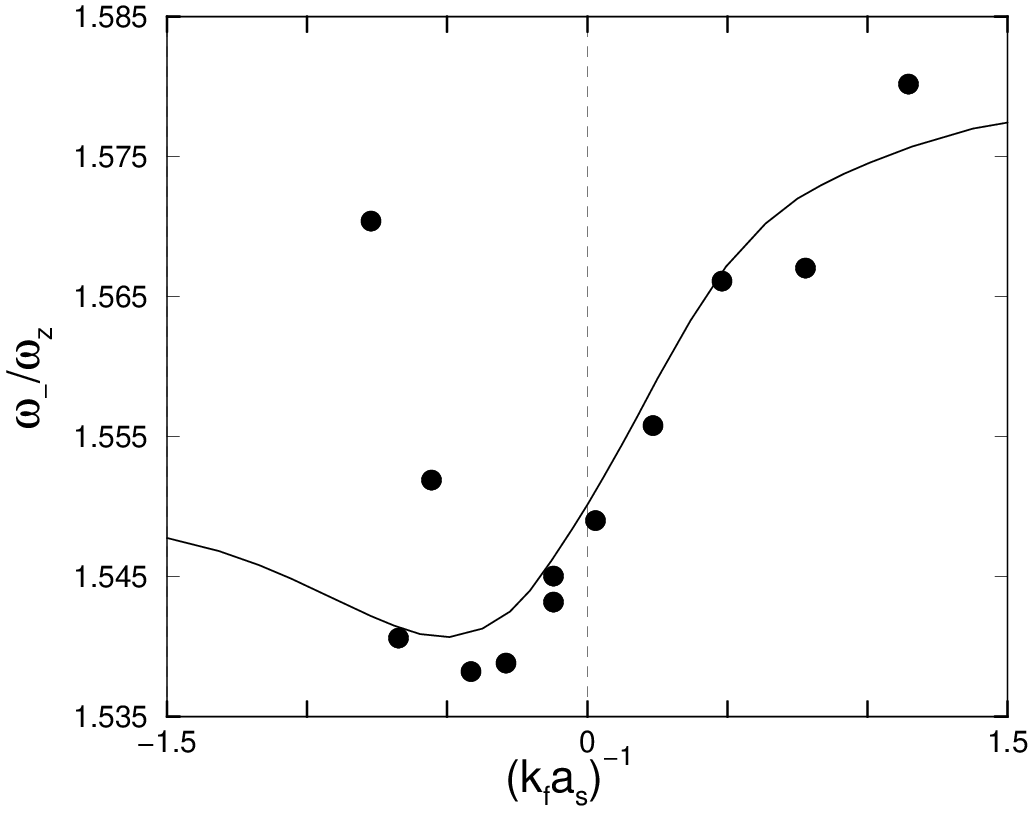}}
\caption{Zero temperature axial breathing mode frequency of a two component Fermi gas in a highly elongated ($\lambda = 0.001$) 
trap within the BCS-BEC crossover 
region. The frequency is given as a function of dimensionless parameter $(k_fa_s)^{-1}$. Notice, at the weak coupling BCS limit 
$[(k_fa_s)^{-1} \rightarrow -\infty]$, 
$\omega_-/\omega_z \rightarrow \sqrt{\frac{12}{5}} $ and at the weakly repulsive BEC limit $[(k_fa_s)^{-1} \rightarrow \infty]$, 
$\omega_-/\omega _z\rightarrow \sqrt{\frac{5}{2}}$ as one expects from hydrodynamic 
theory. At unitarity, $\omega_-/\omega_z \rightarrow \sqrt{\frac{12}{5}}$ as required by universality.
The solid circles are finite temperature experimental results from Ref~\cite{barten1}.}
\end{figure}

\begin{figure}[tbh]
\epsfxsize=8.0cm 
\centerline{\epsffile{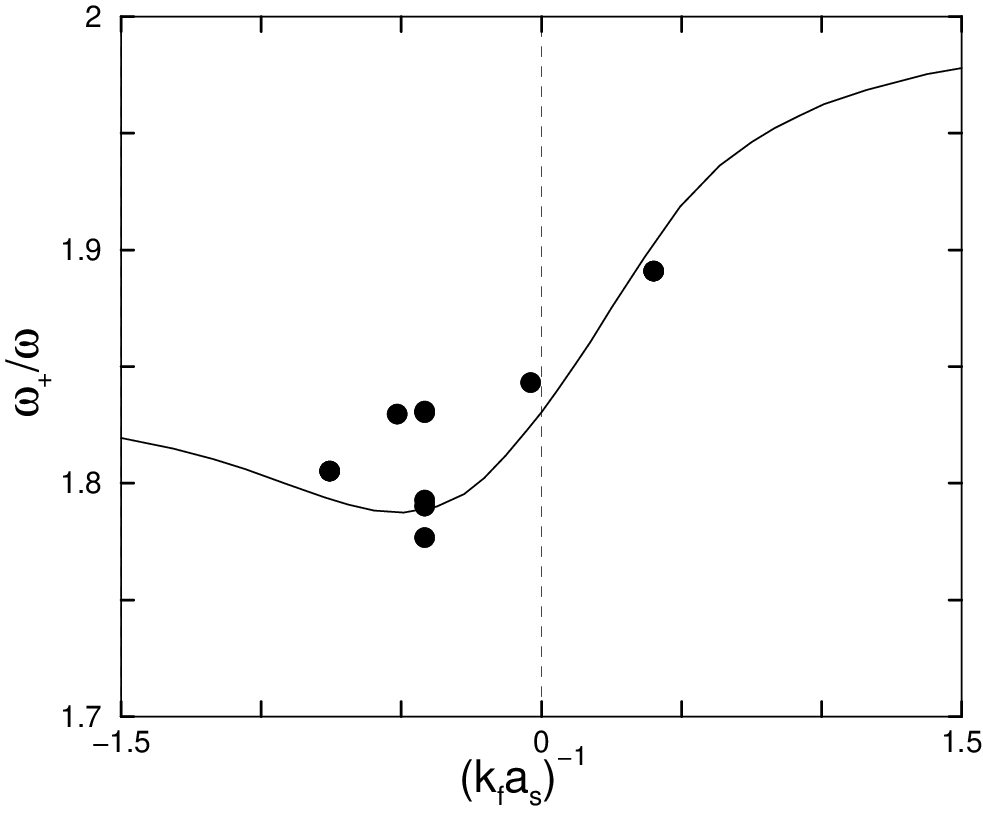}}
\caption{Zero temperature transverse breathing mode frequency of a two component Fermi gas in a highly elongated 
($\lambda = 0.001$) trap within the BCS-BEC crossover 
region. The frequency is given as a function of dimensionless parameter $(k_fa_s)^{-1}$. Notice, at the weak coupling BCS 
limit $[(k_fa_s)^{-1} \rightarrow -\infty]$, 
$\omega_+/\omega_{\perp} \rightarrow \sqrt{\frac{10}{3}}$ and at the weakly repulsive BEC limit $[(k_fa_s)^{-1} \rightarrow \infty]$, 
$\omega_+/\omega_{\perp} \rightarrow 2$ 
as one expects from hydrodynamic 
theory. At unitarity, $\omega_+/\omega_{\perp} \rightarrow \sqrt{\frac{10}{3}}$ as required by universality. 
The solid circles are finite temperature experimental results from Ref~\cite{kinast1}. The multitude of experimental data points 
at $(k_fa_s)^{-1} \sim -0.42$ represent different temperatures.} 
\end{figure}

\section{V. Discussion and conclusions}

Our results are in quantitative agreement with the experiments carried out by Kinast et al~\cite{kinast1} and 
Bartenstein et al~\cite{barten1} on the Fermi gas of $^6Li$ atoms. While transverse breathing modes have been measured in both 
experiments, axial breathing modes have been measured only by  Bartenstein et al~\cite{barten1}. 
The measured finite temperature axial breathing modes 
by Bartenstein et al~\cite{barten1} initially show a decrease in the collective excitation frequency as the magnetic field 
is swept from the weakly repulsive BEC limit towards the BCS limit. Experimentally a dip is 
found at $(k_fa_s)^{-1} \sim -0.5$ after which there is a striking increase in the oscillation frequency (see FIG. 3). 
The location of the dip is in excellent agreement with our 
prediction which shows a similar dip at $(k_fa_s)^{-1} \sim -0.46$. The measurements at the weakly repulsive BEC 
limit ($\omega_- = \sqrt{5/2}\omega_z$) and at unitarity ($\omega_- = \sqrt{12/5}\omega_z$) are in quantitative agreement with 
our prediction. However, the measured collective axial breathing mode frequencies at the weak coupling BCS limit approach 
$2\omega_z$, while, as with most other theories, we predict that $\omega_- \rightarrow \sqrt{12/5}\omega_z$. 
In fact at $(k_fa_s)^{-1} \sim -0.67$, the upper bound calculated from our sum rules begins to be below the 
measured axial mode frequencies. 
As suggested by Bartenstein et al~\cite{barten1}, this jump may indicate a breakdown of the hydrodynamic theory, and is undoubtedly 
related to pair breaking as at this point the mode frequency is comparable to twice the gap. If the experiment is done 
in the ranges of densities and temperatures where the threshold energy of the two-particle continuum $2E_g$ 
(where $E_g = \tilde{\Delta}$  for $\mu > 0$ and $E_g = \sqrt{\mu^2+\tilde{\Delta}^2}$ for $\mu < 0$, is the single 
particle excitation gap) is smaller than the collective energies, then the collective modes can become overdamped, 
leaving the system to simply oscillate at a multiple of the trap frequency. Since we work in LDA and make mean field 
approximations our upper bound is not rigorous in this region.
 
As shown in FIG. 4, our prediction for the transverse breathing mode frequencies are in quantitative 
agreement with the finite temperature experimental measurements by Kinast et al~\cite{kinast1}. The experimental data points 
at $(k_fa_s)^{-1} \sim -0.42$ represent different temperatures. Surprisingly, we have a good agreement with higher 
temperatures at $(k_fa_s)^{-1} \sim -0.42$. Our predictions 
do not, however, agree with the transverse breathing mode data reported in Ref.~\cite{barten1}, which 
show an abrupt change and large damping close to the 
unitary limit. This disagreement may be due to ellipticity in their optical trap~\cite{rudi}.
 
Finally, our results can be compared to the other theoretical approaches. We have very good agreements in the entire 
BCS-BEC crossover region with the theories proposed by Hu et al~\cite{hu} using a scaling ansatz together with a mean field 
description of the BCS-BEC crossover and 
Kim et al~\cite{kim} using an approach based on 
the framework of hydrodynamic theory. However, our predictions are in disagreement in the BEC regime with the findings 
of Stringari~\cite{stringari1}, Heiselberg~\cite{heisel}, and Manini et al~\cite{manini}, 
which predict a maximum in the transverse 
breathing collective mode frequency on the BEC side of resonance. This difference is probably due to our 
neglect of beyond mean field corrections~\cite{dsp,gea}. 

In conclusion, we have used a sum rule approach to study the collective oscillations of a two component trapped 
atomic Fermi gas in the BCS-BEC crossover region. 
This approach allowed us to simply compute the moments of the susceptibility without evaluating the full response function of the 
system. We explicitly treated the molecular bosons and were able to 
retain effective interaction between Feshbach molecular bosons in the entire crossover region. Our calculation 
shows non-monotonic behavior of the 
breathing modes in the BCS-BEC regime. In the weak coupling BCS and BEC limits, our calculated breathing modes 
approach well known hydrodynamic results. At unitarity the breathing modes have same frequencies as those of a non interacting 
Fermi gas, as required by universality. The quadrupole mode frequencies are 
independent of the interactions and the aspect ratio of the trap. Our predictions are in reasonable agreement with 
experiments and other theoretical approaches. 

\section{VI. Acknowledgments}

This work was supported by NSF grant PHY-0456261 and the Alfred P. Sloan Foundation. We are grateful to R. Grimm for sending us his 
data for the axial breathing modes.


\begin{references}
\bibitem{lda1} Validity of the LDA for a Fermi gas requires $n(\vec{r})l_0^4 \gg \vec{r}$, where $l_0$ is 
the oscillator length and $n(\vec{r})$ is the number density at position $\vec{r}$. It is shown that LDA is a reasonably good 
approximation for large N, for an example see Y. Ohashi and A. Griffin, 
Phys. Rev. A 72, 013601 (2005).
\end{references}
\end{document}